
\magnification=\magstep1

\parindent=15pt
\centerline{\hfill OKHEP-93-05}
\bigskip

\centerline{\bf CONTINUED FRACTION AS A DISCRETE NONLINEAR TRANSFORM}
\bigskip
\bigskip
\bigskip
\centerline{Carl M. Bender}
\medskip
\centerline{Department of Physics}
\medskip
\centerline{Washington University}
\medskip
\centerline{St. Louis, MO 63130}
\bigskip
\bigskip
\centerline{Kimball A. Milton}
\medskip
\centerline{Department of Physics and Astronomy}
\medskip
\centerline{University of Oklahoma}
\medskip
\centerline{Norman, OK 73019}
\bigskip
\bigskip
\bigskip
\bigskip
\bigskip
\centerline{\bf ABSTRACT}
\bigskip
The connection between a Taylor series and a continued-fraction
involves a
nonlinear relation between the Taylor coefficients $\{ a_n \}$ and
the
continued-fraction coefficients $\{ b_n \}$. In many instances it
turns out
that this nonlinear relation transforms a complicated sequence $\{
a_n \}$
into a very simple one $\{ b_n \}$. We illustrate this simplification
in the
context of graph combinatorics.
\footnote{}{PACS numbers: 02.90.+p, 11.90.+t, 11.10.-z}
\vfill \eject

The purpose of a transform is to convert an apparently complicated
problem into
one that is obviously simple. In order to be useful a transform must
have
an inverse. One applies the transform to a difficult-looking problem,
solves
the resulting easy problem, and applies the inverse transform to
obtain the
solution to the original problem.

A typical example of a linear transform is the Fourier transform:
$${\cal F}[f]\equiv {1\over{\sqrt{2\pi}}}\int_{-\infty}^{\infty}
dx\;e^{ixy}f(x)
=F(y).\eqno\hbox{(1a)}$$
The inverse transform is defined by:
$${\cal F}^{-1}[F]\equiv{1\over{\sqrt{2\pi}}}
\int_{-\infty}^{\infty}dy\;e^{-ixy}
F(y)=f(x).\eqno\hbox{(1b)}$$
The Fourier transform (1a) converts the heat equation
$$u_t = u_{xx},$$
which is a partial differential equation for $u(x,t)$, into the
ordinary
differential equation
$$U_t = -y^2 U$$
for the function $U(y,t)$. This ordinary differential equation is
easy to
solve and one need only apply the inverse Fourier transform (1b) to
the
solution of the ordinary differential equation to obtain the solution
to the
original heat equation.

One solves the Korteweg-deVries equation, a difficult nonlinear wave
equation,
by means of an interesting {\sl nonlinear} transform, which converts
the
original partial differential equation into a simple linear problem
involving
isospectral flow. The inverse transform is performed by the method of
inverse
scattering.

In this paper we investigate a nonlinear transform that converts the
discrete
sequence $a_1,~a_2,~a_3,~\ldots$ into another sequence
$b_1,~b_2,~b_3,~\ldots$.
The first four equations for this transform are:
$$b_1 = a_1, \eqno\hbox{(2a)}$$
$$b_2 = -a_1 + a_2/a_1, \eqno\hbox{(2b)}$$
$$b_3 = {{a_1 a_3 - a_2^2}\over {a_1 a_2 - a_1^3}},
\eqno\hbox{(2c)}$$
$$b_4 = {{a_1 a_2 a_4 - a_1^3 a_4 - a_1 a_3^2 + 2a_1^2 a_2 a_3 - a_1
a_2^3}
\over{a_1 a_2 a_3 - a_1^3 a_3 - a_2^3 + a_1^2 a_2^2}}.
\eqno\hbox{(2d)}$$
The first four equations for the inverse transform are:
$$a_1 = b_1, \eqno\hbox{(3a)}$$
$$a_2 = b_1 (b_1 + b_2), \eqno\hbox{(3b)}$$
$$a_3 = b_1 [b_2 b_3 + (b_1 + b_2)^2], \eqno\hbox{(3c)}$$
$$a_4 = b_1 [b_2 b_3 (b_3 + b_4) + 2 (b_1 + b_2) b_2 b_3 + (b_1 +
b_2)^3].
\eqno\hbox{(3d)}$$
Observe that the structure of this transform is triangular in the
sense that
the first $n$ terms in the $a$ sequence uniquely determine $b_n$ and
the first
$n$ terms in the $b$ sequence uniquely determine $a_n$.

We can derive the formulas in (2) and (3) for this transformation
very simply.
Consider the formal Taylor series
$$1+ \sum_1^{\infty} (-1)^n a_n x^n\eqno\hbox{(4)}$$
and the continued fraction
$$1/(1+b_1 x/(1+b_2 x/(1+b_3 x/(1 + \cdots .\eqno\hbox{(5)}$$
If we demand that the expressions in (4) and (5) represent the same
function
of $x$ we obtain the relations between the coefficients given in (2)
and (3).

There is an intimate relation between the nonlinear transform in (2)
and (3) and
the theory of orthogonal polynomials.$^{1,2}$ Let the numbers $\{ a_n
\}$
represent the $2n$th moments of some weight function $w(x)$:
$$\int_{-L}^L dx\; x^{2n} w(x) = a_n.\eqno\hbox{(6)}$$
We assume here that $w(x)$ is a positive even function of $x$ so that
the
odd moments vanish and that the numbers $\{ a_n \}$ are all positive.
The value of $L$ is left unspecified. The numbers $\{ b_n \}$ can be
used to
construct a set of polynomials $\{ P_n (x)\}$ by means of the
recursion relation
$$P_{n+1}(x) = x P_n (x) - b_n P_{n-1} (x) \eqno\hbox{(7)}$$
together with the initial conditions $P_0 (x) = 1$ and $P_1 (x) = x$.
Note that the polynomials $\{ P_n (x) \}$ are {\sl monic} (the
coefficient
of the highest power of $x$ is 1). If we demand that the polynomials
$\{ P_n
(x) \}$ be orthogonal with respect to the weight function $w(x)$ this
imposes
a sequence of constraints between the numbers $\{ a_n \}$ and $\{ b_n
\}$.
The constraints are given precisely by the nonlinear relations (2) or
(3).
The normalization of the polynomials $\{ P_n (x)\}$ can be expressed
simply
in terms of the numbers $\{ b_n\}$:
$$\int_{-L}^L dx\; w(x)[P_n(x)]^2 = b_1 b_2 b_3 \cdots
b_n.\eqno\hbox{(8)}$$

Here are some examples for which an apparently complicated sequence
$\{ a_n \}$
transforms into an obviously simple sequence $\{ b_n \}$ under the
nonlinear
transform in (2):
$$a_n = n!, \quad \{ b_n\} =\{
1,~1,~2,~2,~3,~3,~\ldots\},\eqno\hbox{(9a)}$$
$$a_n = (n+1)!, \quad \{ b_n\} =\{
2,~1,~3,~2,~4,~3,~\ldots\},\eqno\hbox{(9b)}$$
$$a_n = (2n-1)!!, \quad b_n =n,\eqno\hbox{(9c)}$$
$$a_n = (3n+1)!!! = \Gamma (n+1/3) 3^n /\Gamma (1/3), \quad \{ b_n\}
=\{ 1,~3,~4,~6,~7,~9,~10,~\ldots\},\eqno\hbox{(9d)}$$
$$a_n = (4n+1)!!!! = \Gamma (n+1/4) 4^n /\Gamma(1/4), \quad \{ b_n\}
=\{ 1,~4,~5,~8,~9,~12,~13,~\ldots\},\eqno\hbox{(9e)}$$
$$a_n = 2^n (2n-1)!!/(n+1)!, \quad b_n = 1,\eqno\hbox{(9f)}$$
$$\{ a_n \} = \{ (-1)^n E_{2n}\} = \{
1,~5,~61,~1385,~50521,~\ldots\},\quad
b_n = n^2,\eqno\hbox{(9g)}$$
where $\{ E_{2n} \}$ are the Euler numbers,
$$\{ a_n \} = \{ (-1)^n 6B_{2n+2}\} = \{
1/5,~1/7,~1/5,~5/11,~691/455,~\ldots\},
\quad b_n = {{n(n+1)^2 (n+2)}\over{4(2n+1)(2n+3)}},\eqno\hbox{(9h)}$$
where $\{ B_{2n}\}$ are the Bernoulli numbers. The simplifying effect
of
the nonlinear transform in (2) is particularly evident for the case
of the Euler
and Bernoulli numbers; there is no simple formula for $a_n$ in (9g)
and (9h).

The examples in (9) illustrate an interesting property of the
nonlinear
transform in (2); if $a_n$ grows like $(kn)!$ for large $n$, then
$b_n$
grows algebraically like $n^k$ as $n\to\infty$. Thus, the asymptotic
growth of $b_n$ is much milder than that of $a_n$.

Most of what we have said so far can be found in various
references.$^3$ The
objective this paper is to show that the nonlinear transform in (2)
and (3) can
be used to simplify the combinatorics of graph counting. There is a
simple
formula for the number of vacuum graphs, connected plus disconnected,
weighted
by their respective symmetry numbers. Specifically, for a $\phi^{2N}$
field
theory, the sum $V_n$ of the symmetry numbers of all of the vacuum
graphs having
$n$ vertices is
$$V_n = {{(2Nn-1)!!}\over {[(2N)!]^n n!}}.\eqno\hbox{(10)}$$
However, the more important quantity in quantum field theory is
$C_n$, the sum
of the symmetry numbers of the $n$-vertex {\sl connected} vacuum
graphs. The
number $C_n$ is directly related to the vacuum energy of the theory.
For
example, for a $\phi^4$ theory,
$$\{ C_n^{[4]} \} =\left\{ {{1}\over{2^3}},~{{1}\over{2^2\cdot
3}},~{{11}
\over{2^5 \cdot 3}},~{{17}\over {2^3\cdot 3^2}},~{{619}\over{2^6\cdot
3 \cdot
5}},~{{709} \over{2^2\cdot 3^4}},~\ldots \right \},\eqno\hbox{(11)}$$
where we have presented the numbers in factored form.

Clearly, there is no simple formula for $C_n$. There is, however, a
nonlinear convolution formula$^4$ that relates $V_n$ and $C_n$:
$$V_n - C_n ={1\over n} \sum_{j=0}^{n-1} jC_j
V_{n-j}.\eqno\hbox{(12)}$$
Notice that in this convolution formula the quantity $C_n$ always
appears
multiplied by $n$. This suggests that we take
$$a_n = 4n C_n^{[4]} 6^n,\eqno\hbox{(13)}$$
where we have inserted the constants 4 and 6 to simplify maximally
the resulting
numbers $\{ b_n \}$. Under the nonlinear transform in (2) we find
that
$$b_n = 2n+1,\eqno\hbox{(14)}$$
a remarkable simplification of the complicated sequence of numbers in
(11).

A similar dramatic simplification occurs in $\phi^3$ theory. For this
theory all
connected vacuum graphs have an even number of vertices. The numbers
$C_{2n}^{[3]}$ form a sequence that does not have a simple formula:
$$\eqalignno{
\{ C_{2n}^{[3]}\} &=\bigg \{ {{5}\over{2^3 \cdot
3}},~{{5}\over{2^4}},~{{5\cdot
13 \cdot 17}\over{2^7 \cdot 3^2}}, ~{{5\cdot 113}\over
{2^7}},~{{5^2\cdot
3313}\over{2^{10}\cdot 3}},\cr
&\qquad {{5^2\cdot 787}\over{2^5 \cdot 3}},~{{5^2\cdot 151
\cdot 479\cdot 709}\over{2^{15}\cdot 3\cdot 7}},~{{5^2\cdot
3229117}\over
{2^{12}}},~\ldots\bigg \}.&(15)\cr}$$
We define the sequence $a_n$ using a formula similar to that in (13):
$$a_n = 6n C_{2n}^{[3]} 4^n.\eqno\hbox{(16)}$$
The corresponding sequence $b_n$ is the same as that in (14) except
that
every third entry is deleted:
$$\{ b_n\} = \{ 5,~7,~11,~13,~17,~19,~23,~25,~\ldots
\},\eqno\hbox{(17)}$$

The sequence $b_n$ in (17) also applies to $\phi^6$ theory. For a
$\phi^6$
theory, the numbers $C_n^{[6]}$ are the same, apart from a factor, as
the
numbers $C_{4n}^{[3]}$ in (15) for a $\phi^3$ theory:
$$C_n^{[6]} = {2\over {30^n}} C_{4n}^{[3]} .\eqno\hbox{(18)}$$
\bigskip
\bigskip

We thank Jacques Perk for useful discussions and the U. S. Department
of Energy
for funding this research.
\vfill\eject

\centerline{\bf REFERENCES}
\bigskip
\item{$^1$} H. S. Wall, {\sl Analytic Theory of Continued Fractions}
(Van Nostrand, New York, 1948), p. 197; W. B. Jones and W. J. Thron,
{\sl
Continued Fractions: Analytic Theory and Applications\/}
(Addison-Wesley,
Reading, MA, 1980), pp. 250-255.
\medskip
\item{$^2$} M. L. Mehta, {\sl Matrix Theory: Selected Topics and
Useful
Results} (Les Editions de Physique, Les Ulis, 1988), pp. 107-109. See
also
C. Itzykson and J. B. Zuber, J. Math. Phys. {\bf 21}, 411 (1980).
\medskip
\item{$^3$} For the results on Euler and Bernoulli numbers see Ref.
2, p. 323.
See also H. Au-Yang and J. Perk, Physica {\bf 144A}, 44 (1987).
\medskip
\item{$^{4}$} C. M. Bender and W. E. Caswell, J. Math. Phys. {\bf
119}, 2579
(1978).
\bye